\documentclass{aa}

\usepackage{graphicx}

\def\emph#1{{\sl #1}}
\newcommand{\ltsima} {$\; \buildrel < \over \sim \;$}
\newcommand{\gtsima} {$\; \buildrel > \over \sim \;$}
\newcommand{\lta} {\lower.5ex\hbox{\ltsima}}
\newcommand{\gta} {\lower.5ex\hbox{\gtsima}}
\usepackage{epsf}

\title{Gravitationally lensed radio emission associated with SMM J16359+6612, 
a multiply imaged submillimeter galaxy behind A2218}

\titlerunning{Lensed radio emission in SMM~J16359+6612}
\authorrunning{Garrett et al.}

\author{M.A. Garrett\inst{1}, K.K. Knudsen\inst{2} \& P.P. van der Werf\inst{3}}

\offprints{garrett@jive.nl}

\institute{
Joint Institute for VLBI in Europe, Postbus 2, 7990 AA, Dwingeloo, The
Netherlands
\and
Max-Planck-Institute f\"ur Astronomie, K\"onigstuhl 17, D-69117
Heidelberg, Germany
\and
University of Leiden, Department of Astronomy, P.O. Box 9513, 2300 RA
Leiden, Netherlands.
}

\date{Received ...; accepted ...}

\begin{document}
\maketitle

\abstract{ 
  
  We report the detection of discrete, lensed radio emission from the
  multiply imaged, $z=2.516$ submillimetre selected galaxy, SMM
  J16359+6612. All three images are detected in deep WSRT 1.4 GHz and
  VLA 8.2 GHz observations, and the radio positions are coincident with
  previous sub-mm SCUBA observations of this system. This is the widest
  separation lens system to be detected in the radio so far, and the
  first time that multiply imaged lensed radio emission has been
  detected from a star forming galaxy --- all previous multiply-lensed
  radio systems being associated with radio-loud AGN. Taking into account the
  total magnification of $\sim 45$, the WSRT 1.4 GHz observations
  suggest a star formation rate of $\sim 500$~M$_{\odot}$~yr$^{-1}$.
  The source has a steep radio spectrum $\alpha \sim -0.7$ and an
  intrinsic flux density of just 3 microJy at 8.2 GHz. Three other
  SCUBA sources in the field are also detected by the WSRT, including
  SMMJ16359+66118, a singly imaged (and magnified) arclet at z=1.034.
  Higher resolution radio observations of SMMJ16359+6612 (and other
  highly magnified star forming galaxies) provide a unique opportunity
  to study the general properties and radio morphology of intrinsically
  faint, distant and obscured star forming galaxies. They can also help
  to constrain the technical specification of next generation radio
  telescopes, such as the Square Kilometre Array.
  

}

\section{Introduction}

SMMJ16359+6612 is a sub-millimetre galaxy (SMG) located at z=2.516 that
is triply imaged and highly magnified by the core of a massive
foreground cluster, Abell 2218 (Kneib et al. 2004a, Knudsen 2004). The
three lensed images of the background source are highly magnified (by
factors of 22, 14 \& 9) and the maximum image separation is $\sim 41$
arcseconds. All three images are detected by the Submillimeter
Common-User Bolometric Array (SCUBA) at 850 \& 450 microns with the
brightest image having a measured 850 micron flux density of 17 mJy
(Kneib et al. 2004a). Recently, both Kneib et al. (2004b) and Sheth et
al. (2004) have observed molecular emission from the CO(3-2) line for
each of the three images. There are two distinct velocity components in
this line separated by 280 kms/s, suggesting a total dynamical mass of
the galaxy of $1.5\times10^{10}$M$_{\odot}$. There is also a spatial
offset of 1 arcsec between these two components and Kneib et al.
(2004b) argue, that the source is likely to be a merging system with
the 2 nuclei separated by 3 kpc.

At radio wavelengths only the bright tail of the SMG population is
usually detectable, so observations of highly magnified systems such as
SMMJ16359+6612, offer a unique opportunity to study the radio
properties of one example of the faint sub-mJy, SMG population.  Little
is known about the nature of these faint SMG, despite the fact that
they dominate (energetically) the cosmic far-infrared background
(Knudsen 2004). The intrinsic (unlensed) sub-mm flux of SMMJ16359+6612
is estimated to be $S_{850}=0.8$ mJy (placing it well below the
confusion limit of normal blank field sub-mm surveys). If one assumes
that the FIR-radio correlation holds for this high-z source (e.g.
Garrett 2002), the intrinsic flux density of any radio counter-part is
estimated to be only a few microJy at 8.2~GHz.  However, the
magnification provided by the lens boosts the flux density of even the
faintest image in SMMJ16359+6612 by a factor of 9, 
suggesting that radio counterparts to the sub-mm sources
associated with SMMJ16359+6612 should be easily detectable in deep
Very Large Array (VLA) and Westerbork Synthesis Radio Telescope (WSRT)
images of the field.  We note that singly imaged SMG, modestly
amplified by foreground clusters, have already been detected at radio
wavelengths e.g. Smail et al. (2000) and Ivison et al. (2001).

In this paper we present WSRT 1.4~GHz and VLA 8.2 GHz observations of
Abell 2218, specifically the area of sky surveyed by the Leiden-SCUBA
lens survey (Knudsen 2004) that includes SMMJ16359+6612. The VLA data
are in the public domain and were extracted from the on-line VLA
archive.  In section 2 we describe the radio observations and data
analysis. The results and discussion are presented in section 3 and we
give our conclusions in section 4. Throughout this paper we assume an
$\Omega=0.3$, $\Lambda=0.7$ cosmology with $H_{\o}=70$
km~s$^{-1}$~Mpc$^{-1}$.

\section{Radio Observations} 

\subsection{WSRT Observations} 

Observations of Abell 2218 at 1.4~GHz were conducted by the WSRT on 5
November 2004. The 12 hour run used the full 160 ($8 \times 20$) MHz
WSRT observing band and employed the default continuum frequency set-up
(with the eight bands centered between 1.311 and 1.450 GHz).
Observations were made assuming a coordinate equinox of 2000.  For each
of the 8 bands, 64 spectral channels were generated (a total of 512
spectral points were obtained for the 160 MHz band) and 4 polarization
products were recorded. Unfortunately, radio telescope ``RT7'' was being
used for single-dish VLBI observations and was unavailable during the
entirety of these observations. A short 20-minute scan on 3C~48 was
used to amplitude calibrate the data. A phase reference source
(J1642+689) was observed for 5 minutes every hour. The data analysis
was performed using the NRAO AIPS package. The fringe-fitted phase
solutions from J1642+689 were applied to the A2218 target field. For
this field, the entire WSRT primary beam was imaged using the AIPS task
IMAGR. Several bright sources were detected in the field of view, and
the data were self-calibrated using these and other sources in the
field as a sky model. These outlying sources were later subtracted from
the data set and self-calibration continued using only those sources
located in the centre of the field. The data were self-calibrated first
in phase and later in both amplitude and phase using the AIPS task
CALIB.

A uniformly weighted (``robust -1'') WSRT image of the field convolved
with a Gaussian restoring beam of $13.6 \times 12.4$ arcseconds (in
position angle, $PA= -0.21^{\circ}$) is shown in Figure~1. At 1.4~GHz
there is a great deal of extended emission in the vicinity of the A2218
cluster core.  In order to enhance the contrast of the WSRT image,
spacings shorter than 800$\lambda$ were not used to form the image
presented in Figure 1. The image reaches a 1 $\sigma$ rms noise level
of $\sim 15 ~\mu$Jy~beam$^{-1}$.

   \begin{figure}[ht]
   \centering
   \includegraphics[scale=0.50,angle=0]{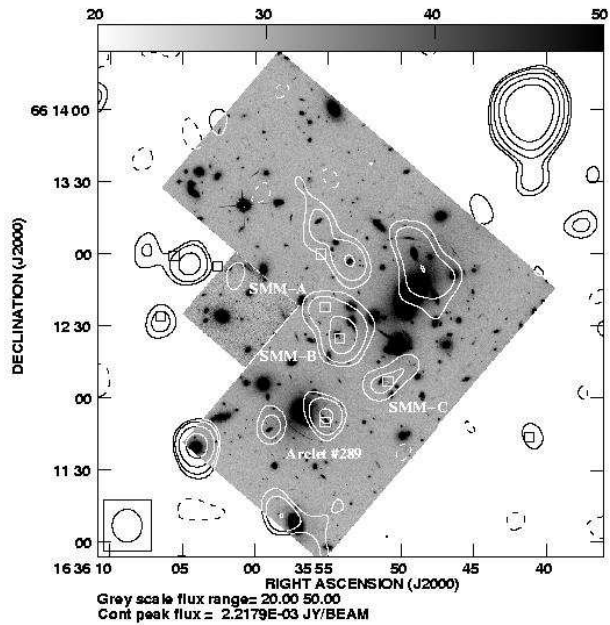}
   \caption{
     The WSRT 1.4 GHz contour map superimposed upon an HST F702W image
     of the region of sky associated with SMM~J16359+6612. Radio
     emission from all three lensed images (SMM-A, SMM-B and SMM-C) is
     detected by the WSRT. The boxes represent the SCUBA source
     positions. Radio emission from three other SCUBA sources in the
     field (including the highly magnified arclet \#289,
     SMM~J16359+66118) are also boxed. Contours are drawn at
     -3,3,5,10,20 \& 40 times the 1-$\sigma$ noise level of 15
     microJy/beam.}
              \label{Fig1}
    \end{figure}

\subsection{VLA observations} 

We searched the VLA archive for observations associated with the
central area of Abell 2218. Several data sets were identified that
included short, snapshot observations of this field. Only one 
full-track 8.2 GHz observation of A2218 was identified, {\it AR045}.
These X-band observations were made on 1999 June 18 UT with the VLA in
its D (3 km) configuration. Data were acquired in dual circular
polarizations with bandwidths of 25 MHz and at center frequencies of
8.1732, and 8.2732 GHz. Each of these 25 MHz IFs was further divided
into seven spectral channels with a width of 3.125 MHz.  Observations
were made assuming a coordinate equinox of 2000. The absolute flux
density scale was set by observations of 3C 286. The data calibration
was made in the standard way with the phase variations
during the observations calibrated via short observations of the VLA
calibrator source 1642+689. Unfortunately observations of 1642+689 were
only made 3 times over the course of the 24 hour observations. The
calibrated A2218 data were clipped in order to remove a few high points
before beginning the self-calibration process. The self-calibration
process followed the same outline as that described for the WSRT
observations.

A naturally weighted VLA image of the field convolved with a Gaussian
restoring beam of $11.2 \times 10.4$ arcseconds (in position angle,
$PA= -72^{\circ}$) is shown in Figure~2.  The image has an r.m.s. noise
level of 6 microJy beam. 

   \begin{figure}[ht]
   \centering
   \includegraphics[scale=0.50,angle=0]{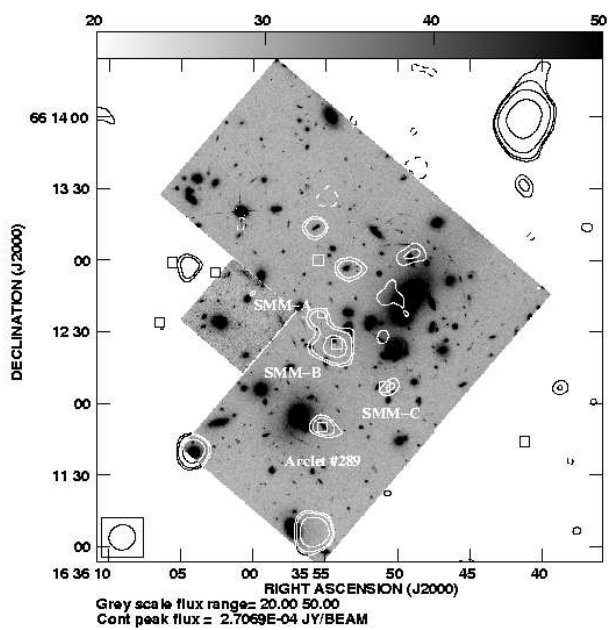}
   \caption{
     The VLA 8.2 GHz contour map superimposed upon an HST F702W image
     of the region of sky associated with SMM~J16359+6612F702W. Radio
     emission from all three lensed images is detected by the WSRT.
     The boxes represent the SCUBA source positions. Radio emission
     from the highly magnified arclet (\#289), SMM~J16359+66118 is also
     detected. Contours are drawn at -3,3,4,7 \& 10 times the
     1-$\sigma$ noise level of 6 microJy/beam.}
              \label{Fig2}
    \end{figure}

\section{Results \& Discussion} 

Figures 1 and 2 present the uniformly weighted WSRT and naturally
weighted VLA contour maps of part of the A2218 field, superimposed upon an
HST F702W image (Kneib et al. 1996) of the region of sky that includes
the lensed sub-mm galaxy SMM~J16359+6612 (Kneib et al. 2004a). The
positions of the nine sources detected by SCUBA at 850 micron (Knudsen
2004) are also identified as square boxes in these figures.  Following
Kneib et al.  (2004a) we identify the lensed images as SMM-A
(SMMJ16359+6612.6), SMM-B (SMMJ16359+6612.4) and SMM-C
(SMMJ16358+6612.1). Both the WSRT and VLA observations detect radio
emission associated with all three lensed images. The faintest SCUBA
detection (SMM-C) is detected at the 4-sigma level by the VLA at 8.2
GHz and at the 7-sigma level by the WSRT at 1.4 GHz.

\begin{table*}
\centering
\caption{Details of the WSRT and VLA radio sources associated with
  SMM~J16359+6612 (SMM-A, B, C) and SMM~J16359+66118 (arclet \#289).}
\begin{tabular}{lccccccc}
\hline\hline\\
Name & RA ($+ 16^{h}$ $35^{m}$) & DEC ($+ 66^{\circ}$) & S$_{T}$ & S$_{Pk}$ & Maj
Axis & Min Axis & PA \\
     & J2000 (sec)  & J2000 ($\prime$, $\prime\prime$)& $\mu$Jy & $\mu$Jy & $\prime\prime$ &
     $\prime\prime$ & deg           \\
\hline \\
WSRT~J163550+661205 (SMM-C) & $51.1\pm0.1$ &  12, $08.0\pm1.0$ &
$110\pm17$  & $110\pm17$ & - & - & - \\ 
WSRT~J163553+661226 (SMM-B) & $53.9\pm0.1$ & 12, $26.4\pm0.6$ & $391\pm43$
& $217\pm16$ & $15\pm2$ & $9\pm2$  
& $165\pm12$  \\
WSRT~J163555+661237 (SMM-A) & $55.6\pm0.2$ & 12, $37.7\pm1.0$ & $99\pm17$ & $99\pm17$ & - &
- & - \\ 
VLA~J163550+661206 (SMM-C) &  $50.5\pm0.2$ & 12, $06.8\pm0.1$ &
$24\pm10$ & $24\pm6$ & - & - &
- \\ 
VLA~J163554+661223 (SMM-B) &  $54.5\pm0.1$ & 12, $25.3\pm1.1$ & $124\pm18$
& $51\pm6$ & $16\pm3$ & $10\pm3$
& $39\pm12$ \\
VLA~J163555+661236 (SMM-A) &  $55.7\pm0.1$ & 12,  $37.1\pm0.5$ & $28\pm6$ & $28\pm6$ & - & - & - \\ 
\hline
WSRT~J163555+661152 (\#289) & $55.2\pm0.1$ & 11, $52.9\pm0.6$ &
$170\pm29$ & $175\pm17$ & $6\pm4$ & - & $43\pm20$ \\ 
VLA~J163555+661150 (\#289) & $55.1\pm0.2$ & 11, $50.1\pm0.6$ & $35\pm10$
& $37\pm6$ & $8\pm3$ & - & $65\pm10$ \\  
\hline 
\end{tabular}
\end{table*}

The AIPS task IMFIT was used to fit Gaussian components to the WSRT and
VLA radio sources.  Details of the radio source's total flux density
(S$_{T}$), peak flux density (S$_{Pk}$), position and (where
appropriate) deconvolved Gaussian sizes (major axis, minor axis and
position angle) are detailed in Table 1, together with their associated
formal errors. It was not possible to robustly constrain the
deconvolved size of the Gaussian component associated with the fainter
images SMM-A and SMM-C in the VLA and WSRT images. For SMM-A and SMM-C
the Guassian component size was thus fixed to the respective restoring
beam sizes.  For both the VLA and WSRT images, two Gaussians were fitted
to the SMM-A,B region simultaneously. 

The SCUBA, VLA and WSRT observations have comparable resolution. The
error in the SCUBA 850 micron positions (including confusion effects)
is expected to be $\sim 4$ arcseconds (Knudsen 2004). The radio
positions are expected to be better than this, with an accuracy of
2 arcsecond or better. Slightly larger errors are possible for the
radio positions derived by the WSRT, due to mild confusion in the
field.  The radio and sub-mm positions of SMM-A,B,C are all
consistent within the errors - the largest deviation is an offset of
3 arcseconds (the VLA/SCUBA comparison in the case of the faintest
image SMM-A). There are no obvious systematic offsets in the sub-mm and
radio positions.

At 8.2~GHz the total flux density of SMM-B is 124 microJy, somewhat
larger than our original expectations (see section 1). By co-adding the
total flux density of all three images at 1.4 and 8.2 GHz, we derive a
spectral index of the source $\alpha=-0.7$, similar to the steep values
measured by Richards (2000) for star forming galaxies in the Hubble
Deep Field North. Kneib et al. (2004b) have shown that the overall SED
of SMMJ16359+6612 is similar to Arp~220. By adopting an Arp~220 SED, we
estimate the source's k-corrected radio luminosity at 1.4 GHz to be
$\sim 4 \times 10^{23}$~Watts/Hz. Using the relation between Star
formation rate (SFR) and radio luminosity (Condon 1992), we derive an
intrinsic star formation rate (SFR) for SMM~J16359+6612 of
500~M$_{\odot}$~yr$^{-1}$. This is in good agreement with the value
obtained by Kneib et al. (2004) from the FIR luminosity. As noted by
Kneib et al. the uncorrected H$\alpha$ SFR estimate is only
11~M$_{\odot}$~yr$^{-1}$, suggesting that this galaxy is highly
obscured by dust. 

The VLA 8.2 GHz observations best resolve the lensed images. The flux
density ratio of the images at 8.2 GHz (and their formal error) are
$\sim 0.23\pm0.06$ and $\sim 0.19\pm0.08$ for SMM-A/SMM-B and
SMM-C/SMM-B respectively. Similarly, the WSRT 1.4 GHz flux image
density ratios are $0.25\pm0.05$ and $0.28\pm0.06$ respectively.  The
radio flux density measurements at 1.4 and 8.2 GHz are thus consistent
with the source being gravitationally lensed.  In comparison, the SCUBA
observations give higher values of 0.64 and 0.53, the IRAM CO(3-2) line
intensities (Kneib et al.  2004b) give 0.67 and 0.63, and the K-band
observations (Kneib et al.  2004a) give 0.53 and 0.4. Lens models of
this system (based on Kneib et al. 1996), predict flux density ratios
of 0.63 and 0.4. Discrepancies in the flux density ratio at different
wavelengths might be explained if the various emission regions are not
co-located or have very different size-scales over which the
magnification may change. We also note that the flux density ratio of
SMM-A, B and C between 850~micron and 1.4 GHz are $111\pm22$, $43\pm7$
and $82\pm16$. By comparison the 850~micron/1.4 GHz flux density ratio
of another SMG also located at $z\sim 2.5$, SMM~J14011+0252 (Ivison et
al. 2000) is $127\pm37$. It may be that in the case of SMM~J16359+6612,
the discordant flux ratios at both 1.4 and 8.2 GHz can be explained by
enhanced radio emission associated with SMM-B, perhaps from another
radio source also located in this region of the field.  Higher resolution
radio observations are required in order to settle this question. 

As shown in Table 1, the brightest image SMM-B is resolved by the VLA
8.2 GHz observations. The measured PA of the major axis is $\sim 39$
degrees, in good agreement with the position angle of the major axis of
the CO(3-2) measurements of the same component, $\sim PA=30$ degrees.
Both measurements are consistent with the overall extension of the
associated arc-like HST images (Kneib et al. 2004a). The FWHM of the
major axis of the SMM-B (see Table 1) does not exceed $\sim 17$
arcseconds.  With a magnification of $\sim 22$ for this image, the
intrinsic size must be on the arcsecond or sub-arcsecond scale -
consistent with the sub-galactic sizes measured for radio sources in
the HDF-N.

The WSRT observations also detect other radio counterparts to other
sub-mm sources detected by SCUBA in this field (Knudsen 2004). In
particular, we detect radio counter-parts to SMMJ16357+66117 ($4\sigma$
detection), SMM~J16361+66126 \& SMM~J16359+66118. The latter source
(see Table 1) is previously identified by Kneib et al. (1996) as a
z=1.034 singly-imaged lensed arc (also known as arc \#289, see Swinbank
et al. 2003 and references therein) with a magnification of $\sim 7$.
The source is also detected by the 8.2 GHz VLA observations presented
here, and by the ISOCAM at 15 micron (Barvainis, Antonucci \& Helou
1999).  In both the WSRT and VLA images, the measured size of the
arclet is less than the restoring beam, in particular the minor axes of
the fitted Gaussians are unresolved. The position angle of the major
axes are consistent with the optical extension of this system. This
source also has a steep spectral index, $\alpha \sim 0.9$. The VLA
8.2~GHz observations do not detect the other SCUBA sources -
SMMJ16357+66117 \& SMM~J16361+66126. These are fainter than
SMM~J16359+66118, and must have fairly steep spectral indices ($\alpha
> 0.5$). Further details of the radio counterparts to these SCUBA
sources will be reported elsewhere.

\section{Conclusions}

We have detected with the WSRT at 1.4~GHz and VLA at 8.2~GHz, radio
emission associated with the triply lensed SMG, SMMJ16359+6612.  This
is the first time that radio emission has been detected in a multiply
imaged SMG system lensed by a foreground cluster. The maximum image
separation is $\sim 41$~arcseconds, much larger than any other lens
system detected in the radio. This is also the first time that multiply
lensed radio emission has been detected from a star forming galaxy ---
all previous multiply imaged radio lensed systems are associated with
radio-loud AGN. The properties of the three radio sources are largely
consistent with the gravitational lensing hypothesis. In addition to
detecting SMMJ16359+6612, we also detect radio emission coincident with
SMMJ16359+66118 a singly imaged arclet (\#289).  The sources are only
detectable by current radio instruments due to the high magnification
factor provided by the lens ($\sim 45$ in the case of SMMJ16359+6612).
The intrinsic total flux density of the radio source at 1.4 and 8.2 GHz
is $\sim 14$ and $\sim3$ microJy.

Follow-up radio observations of SMMJ16359+6612 with much better angular
resolution than those presented here are warranted.  Since
sub-arcsecond imaging is routinely possible at radio wavelengths, it
should be possible to fully capitalise on A2218 as a ``natural
telescope''. A comparison with the detailed HST images of this system
will be particularly interesting. Measurements of the size and a study
of the morphology of the radio emitting region should also constrain
the extent of the star formation in this galaxy. Upgraded radio
telescopes (such as the EVLA and e-MERLIN) should be able to
detect many more highly magnified lens systems lying behind massive
foreground clusters.  In very deep integrations, the e-MERLIN and the
EVLA telescopes now under development, may be able to detect radio
sources associated with individual SNe (or SNR), assuming these have
luminosities similar to those detected in local mergers, such as Arp
220. Radio observations of SMMJ16359+6612 and other highly magnified
SMG using existing and upgraded radio telescope facilities, will be
important in determining the general radio properties (e.g.  angular
size) of the faint SMG population, providing essential input to the
technical specification of next generation instruments, such as the
Square Kilometre Array.

\begin{acknowledgements}
  We would like to thank the staff of the WSRT including Rene
  Vermeulen, Raffaella Morganti \& Willem Baan who helped make these
  observations possible at very short notice. We also thank the
  referee, J.-P. Kneib for useful comments and suggestions, that have
  improved the paper. The WSRT is operated by ASTRON (The Netherlands
  Foundation for Research in Astronomy) with support from the
  Netherlands Foundation for Scientific Research (NWO). The National
  Radio Astronomy Observatory is a facility of the National Science
  Foundation operated under cooperative agreement by Associated
  Universities, Inc. This work was supported in part by the European
  Communitys's Sixth Framework Marie Curie Research Training Network
  Programme, Contract No.  MRTN-CT-2004-505183 ``ANGLES''.

\end{acknowledgements}

\end{document}